\documentclass[twocolumn,aps,superscriptaddress,showpacs,nofootinbib]{revtex4-2}
\usepackage{hyperref}
\usepackage{footnote}
\usepackage{amssymb}
\usepackage{amsmath}
\usepackage{graphicx}
\usepackage[normalem]{ulem}
\usepackage{url}
\usepackage{csquotes}
\usepackage{color}
\usepackage{hyperref}

\begin{document}

\title{Constraining the density dependence of the symmetry energy: the isospin transport ratio revisited}

\author{S. Mallik}
\email{swagato@vecc.gov.in}
\affiliation{LPC Caen IN2P3-CNRS/EnsiCaen et Universite, Caen, France}
\affiliation{Physics Group, Variable Energy Cyclotron Centre, 1/AF Bidhan Nagar, Kolkata 700064, India}
\author{F. Gulminelli}
\affiliation{LPC Caen IN2P3-CNRS/EnsiCaen et Universite, Caen, France}

\begin{abstract}
The isospin diffusion of the quasi-projectile formed in the $^{64,58}Ni$ on $^{64,58}Ni$ reactions in the Fermi energy domain is investigated in the framework of the Boltzmann-Uehling-Uhlenbeck transport model.
The well known isospin transport ratio observable is revisited, with the aim of insuring an optimal comparison between experimental data and theoretical calculations and reducing the present uncertainties in the extraction of empirical equation of state parameters.
We show that isospin transport ratios are sensitive to all the low order isovector parameters ($E_{sym}$, $L_{sym}$ and $K_{sym}$).  We demonstrate that realistic models of the equation of state, covering the uncertainty that presently affects the theoretical description of neutron stars static observables, can be effectively discriminated by isospin diffusion experiments, provided the neutron to proton ratio of the projectile remnant is precisely measured as a function of centrality.

\end{abstract}

\maketitle
\section{Introduction}
The recent observation of gravitational waves (GW) from binaries of compact stars \cite{ligo_O1,ligo_O3,ligo_O3bis}
has greatly renewed the interest towards the nuclear matter equation of state (EoS), which is the unique physical ingredient needed to theoretically predict all the static observables of neutron stars. In this context, a controlled modeling of the EoS under the hypothesis of a purely nucleonic contribution would act as a null hypothesis allowing to extract from the astrophysical data fundamental information such as the presence of deconfined matter in the core of compact stars and/or the reliability of theories of alternative gravity, see ref.\cite{Baiotti19} for a recent review.

For this ambitious goal, ab-initio calculations of the EoS \cite{Carlson2015} are not sufficient, and constraints are needed from laboratory experiments, see  ref.\cite{Burgio2020} for a review. Isospin transport in heavy ion collisions is one of the first observables proposed in the literature \cite{Rami, Danielewicz2, Bao-an-li4, Tsang2004} to pin down the density dependence of the  EoS of neutron rich matter, and the seminal work by the MSU experimental group lead to the first compilations of EoS constraints from laboratory experiments \cite{Tsang2012}. Indeed, heavy ion reactions 
provide a unique opportunity to study the matter behavior at both supra-saturation and sub-saturation densities \cite{Danielewicz,Baran, Bao-an-li3,Tsang_rep,horowitz2014,russotto2016,Camaiani}.
Relativistic energy heavy ion (HI) experiments are obviously ideal probes of the high density domain well above the saturation density $\rho_0$, which is the most relevant to compact star physics\cite{russotto2016,Tsang2019,hillman2020}, but
low energy HI reactions are also important and can give complementary information, because uncertainties on the elementary reaction rates are strongly reduced below the pion production threshold.  If tight contraints were obtained on the low order empirical parameters, this would greatly improve the reliability of the extrapolation to neutron star densities. In particular,  it was shown in ref.\cite{tews18} that a 60\% error bar on the pressure at $2\rho_0$ would be enough to produce constraining predictions for the tidal polarizability extracted from GW measurements, and potentially discriminate between EoS with and without transitions to deconfined matter.

The limitation of the present experimental constraints in the Fermi energy regime, is the uncertainty linked to the comparison protocol between nuclear experiments and transport models \cite{coupland2011}. The pioneering work by Tsang et al. \cite{Tsang2000} clearly showed that the isospin transport ratio is very well correlated to the density dependence of the symmetry energy, and allowed excluding extreme values for the $E_{sym}$ and $L_{sym}$ parameters \cite{Tsang2012}. However  tighter constraints could not be obtained, mainly because the one-body observable which is most robustly predicted by the transport model calculations, namely the isotopic ratio of the quasi-projectile remnant, was not directly measurable at that time.  Surrogate variables were therefore employed, such as isoscaling parameters or light cluster isobaric ratios \cite{Tsang2004,Tsang2009,Zhang2020}.
However, as already observed in ref.\cite{coupland2011} and pointed out by the present study, isotopic ratios built out of different isospin sensitive observables are not identical, which leads to a systematic error in the extraction of the EoS parameters.

Coalescence invariant n-p ratios extracted from central $^{124,124}Sn$ and $^{112,112}Sn$ collisions at 110AMeV at the NSCL \cite{coupland2016} have been recently proposed for a more controlled comparison with transport models \cite{morfouace2019}.
A complementary approach is followed by the INDRA/FAZIA collaboration, which has recently measured at GANIL  $^{64,58}Ni$ on $^{64,58}Ni$ reactions at 32A MeV and 52A MeV\cite{manip}, with an experimental apparatus able to simultaneously measure the mass and charge of quasi-projectile (QP) residues on an event-by-event basis \cite{fazia}. The incident energy is too low for the momentum dependence to play an important role, and the symmetry energy is expected to be disentangled from the effective mass splitting.
To determine the precision needed from the measurements in order to discriminate between realistic EoS and
waiting for the availability of those data, it is important to revisit the theoretical predictions on the isospin diffusion ratio on the different observables that can be robustly be compared to experimental data, and their sensitivity to the symmetry energy.

To this aim, in this article isospin transport ratios are investigated from neutron to proton ratio of QP as well as that of free nucleons emitted from the QP independently.  BUU@VECC-McGill transport model is employed for this study  \cite{Dasgupta_BUU1,Mallik9, Mallik10, Mallik14,Mallik22}. A meta-functional for the nuclear equation \cite{Margueron2018a} of state is implemented, which allows reproducing a large set of non-relativistic as well as relativistic energy functionals, together with possible novel density dependences not yet explored in existing functionals.
\\
\indent
\section{Model description}
Since we will focus on the properties of the positive rapidity region, calculations are performed in the projectile frame.
Projectile and target nuclei ground states are built  with a variational method \cite{Mallik9,Lee,Cecil} using Myers density profiles \cite{Myers}. The test particles of isospin $q=p,n$ move in a mean-field 
potential $U_q(\rho_p(\vec{r}),\rho_n(\vec{r}))=U_q^{bulk}(\rho_p(\vec{r}),\rho_n(\vec{r}))+U_q^{surf}(\rho_p(\vec{r}))+\frac{1}{2}(1-\tau_q)U^{coul}$, where $U_q^{bulk}$ represents the bulk part which is calculated from a recently proposed meta-functional \cite{Margueron2018a} based on a polynomial expansion in density around saturation and including deviations from the parabolic isospin dependence through the effective mass splitting in the kinetic term. The finite range term $U_q^{surf}$
\cite{Lenk} does not affect nuclear matter properties but produces  realistic diffuse surfaces, and $U^{coul}$ is standard Coulomb interaction potential with $\tau_q$=-1 (1) for protons (neutrons).  The mean-field propagation is done using the lattice Hamiltonian method which conserves energy and momentum very accurately \cite{Lenk}. Two body collisions are calculated as in Appendix B of ref. \cite{Dasgupta_book} with isospin dependent cross-sections \cite{Cugnon}. Comparison of various observables from BUU@VECC-McGull model with other BUU and quantum molecular dynamics based models can be found in \cite{Zhang_code_comparison,Ono_code_comparison}
\\
\begin{figure}
\begin{center}
\includegraphics[width=\columnwidth]{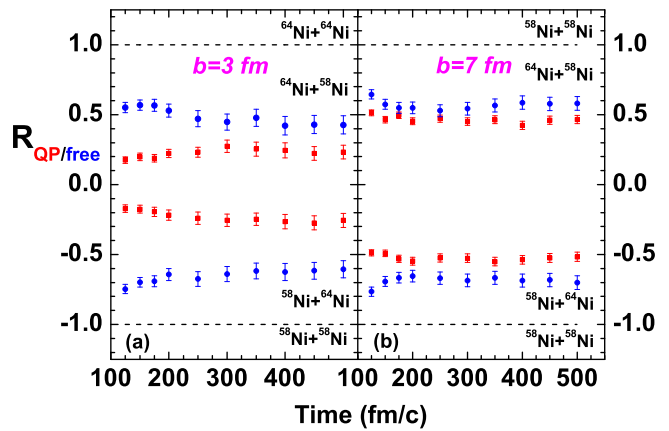}
\caption{Variation of isospin transport ratio calculated from neutron to proton ratio of quasiprojectile (red squares) and free nucleons (blue circles) with time for $^{58,64}$Ni on $^{58,64}$Ni reactions at 52A MeV at impact parameter $b=3$ fm (left panel) and $b=7$ fm (right panel).}
\label{Isospin_imbalance_ratio_SLY5_b=3and7fm_time_dependence}
\end{center}
\end{figure}
\begin{figure}
\begin{center}
\includegraphics[width=\columnwidth]{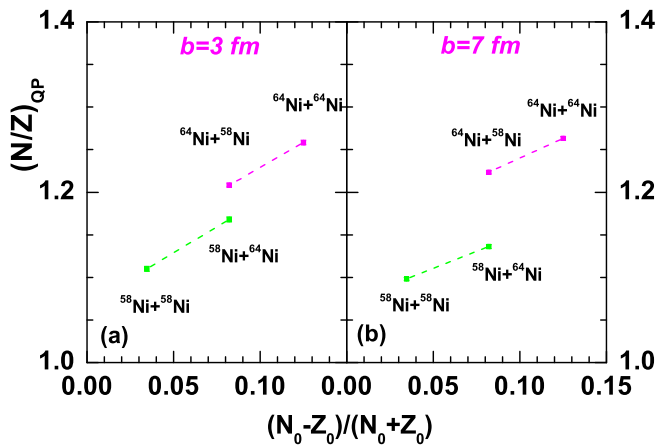}
\caption{Neutron to proton ratio of quasiprojectile as a function of $(N_0-Z_0)/(N_0+Z_0)$ for $^{58,64}$Ni on $^{58,64}$Ni reaction at 52 AMeV at an impact parameter $b=3$ fm (left panel) and $b=7$ fm (right panel). Green squares represent the result with $^{58}$Ni projectile where as magenta squares are for $^{64}$Ni projectile. Dashed lines are drawn to guide the eyes.}
\label{PLF_NbyZ_b=3and7fm}
\end{center}
\end{figure}
\section{Results}
We consider  $^{64,58}Ni$ on $^{64,58}Ni$ reactions at 52A MeV ranging from mid-central collision to very peripheral collision and use as a fiducial EoS the Sly5 functional in BUU@VECC-McGill model. We use 100 test particles per nucleon, and for each reaction 500 events are simulated.  The QP is identified by its momentum distribution peaked at a value close to zero. 
The freeze-out time of the QP is investigated by studying the isotropy of momentum distribution in the QP frame as a function of time \cite{Mallik22,Mallik18}. It is observed that 
isotropy of momentum distribution is achieved at around 100 $fm/c$, and this freeze-out time is almost independent of impact parameter as well as projectile target combination. It is customary in heavy ion reactions to stop the dynamical evolution at the freeze-out time and to couple the transport model to a statistical evaporation model \cite{Camaiani,Bao-an-li2,Mallik11} which allows  determining cross-sections of free nucleons as well as small clusters. But substantial error may arise due to (i) uncertainty of precise coupling time and excitation and (ii) inconsistency between the EoS used in the transport model and binding as well as level density used in the afterburner. To overcome this difficulty, we do not couple the dynamical model to a statistical evaporation model and rather   study how  the observables of  interest do depend on time. The most sensitive observable to isospin diffusion is the isospin transport ratio, for which diffusion free limits are given by the reactions with identical projectiles and targets i.e. $^{58}Ni$ on $^{58}Ni$ and $^{64}Ni$ on $^{64}Ni$ reactions. On the other side, for $^{58}Ni$ on $^{64}Ni$ and $^{64}Ni$ on $^{58}Ni$ reactions, there will be opportunity of isospin diffusion from the more isospin asymmetric system to the more symmetric one,  via nucleon transfer through the overlap region. Following refs. \cite{Rami,Tsang2004,Tsang2009}, the isospin transport ratio is expressed as
\begin{equation}
R=\frac{2x_{^{A_p}Ni+^{A_t}Ni}-x_{^{58}Ni+^{58}Ni}-x_{^{64}Ni+^{64}Ni}}{x_{^{58}Ni+^{58}Ni}-x_{^{64}Ni+^{64}Ni}}\nonumber\\
\end{equation}
where $x_{^{A_p}Ni+^{A_t}Ni}$ is an isospin sensitive observable which can be calculated for each reaction between projectile and target mass number  $A_p$ and $A_t$ respectively.\\
\begin{figure}
\begin{center}
\includegraphics[width=\columnwidth]{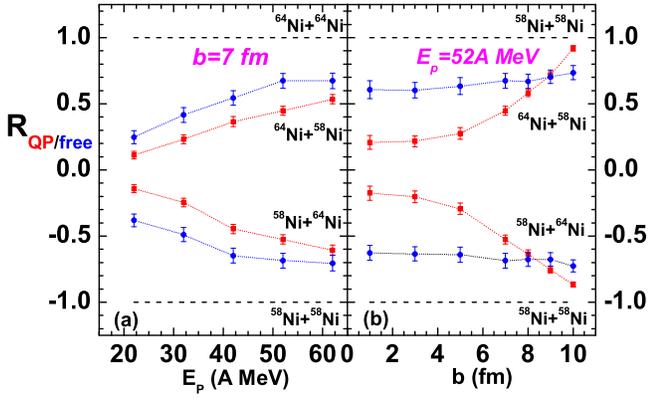}
\caption{Projectile energy (left panel) and impact parameter (right panel) of isospin transport ratio calculated from neutron to proton ratio of quasiprojectile (red squares) and free nucleons (blue circles) with time for $^{58,64}$Ni on $^{58,64}$Ni. Projectile energy ($E_p$) dependence is studied at constant impact parameter 7 fm where as for centrality dependence projectile energy is fixed at 52A MeV. Dashed lines are drawn to guide the eyes.}
\label{Isospin_imbalance_ratio_SLY5_energy_and_centrality_dependence}
\end{center}
\end{figure}
\begin{figure}
\begin{center}
\includegraphics[width=\columnwidth]{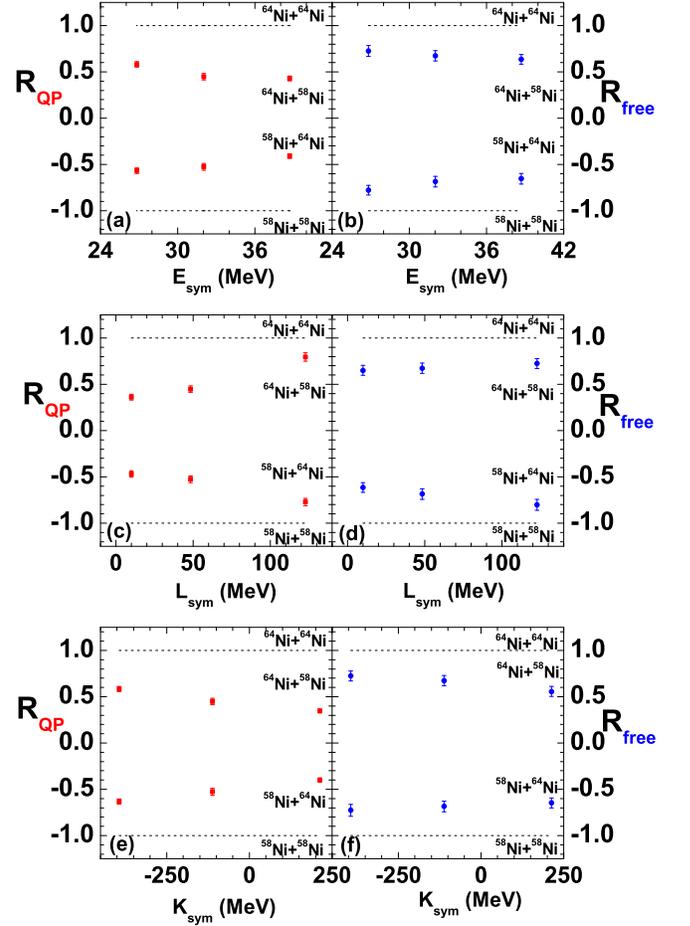}
\caption{Sensitivity of isospin imbalance ratio calculated from neutron to proton ratio of quasiprojectile (left panels) and free nucleons with $p_zc>0$ MeV (right panels) with $E_{sym}$ (top panels), $L_{sym}$ (middle panels) and $K_{sym}$ (bottom panels) for $^{58,64}$Ni on $^{58,64}$Ni reaction with projectile beam energy 52A MeV at impact parameter $b$=7 fm.}
\label{Isospin_imbalance_ratio_sensitivity}
\end{center}
\end{figure}
\indent
By construction, $R$ is 1 for the $A_p=A_t=64$ reaction and -1 for $A_p=A_t=58$ reaction.
In the transport modeling, two different correlated observables can be built, both
linked to the isospin properties of the projectile remnant, (i)$x=N/Z$ of the QP (which gives $R_{QP}$) and (ii) $x=N/Z$ of the free nucleons emitted from the QP (which gives $R_{free}$). The first observable is a robust prediction of mean-field models. In particular, it is only marginally affected by the dynamics in the neck region and the associated cluster production, which is not satisfactorily treated in transport theories. Concerning the second variable,
in order to reduce the contribution of nucleons originated from the neck region, a constant momentum cut at $p_zc=0$ MeV is applied. Fig. \ref{Isospin_imbalance_ratio_SLY5_b=3and7fm_time_dependence}  shows the variation of $R_{QP}$ from  freeze-out to a very large time which can be considered as asymptotic time. It can be seen that $R_{QP}$ saturates around t= 300 $fm/c$  , showing that this observable is not affected by secondary decay. The same is approximately true for $R_{free}$. We therefore choose t= 300 $fm/c$ for the subsequent analysis.\\
\indent
The isospin diffusion effect is shown in Fig. \ref{PLF_NbyZ_b=3and7fm} for two different impact parameters. The
increased (decreased) value of $(N/Z)_{QP}$ for $^{58}Ni$ on $^{64}Ni$ ( $^{64}Ni$ on $^{58}Ni$) with respect to  $^{58}Ni$ on $^{58}Ni$  ( $^{64}Ni$ on $^{64}Ni$) signals the isospin diffusion phenomenon, which tends to equilibrate the global $N/Z$ ratio, and is more effective for central reactions corresponding to increased overlap between the collision partners.
A similar behavior is observed for $(N/Z)_{free}$ (not shown).
\indent
Fig. \ref{Isospin_imbalance_ratio_SLY5_energy_and_centrality_dependence} represents the entrance channel effect on isospin diffusion. Left panel shows the beam energy dependence at constant impact parameter $b=$7 $fm$ whereas the right panel represents the impact parameter variation at constant  beam energy 52A MeV. Isospin transport ratios calculated from both $N/Z$ of QP as well as that of free nucleons are shown in the same diagram.
The qualitative behavior of $R_{QP}$ is in agreement with previous studies \cite{Tsang2004,Tsang2009,Zhang2020}.
The very weak dependence on impact parameter of the free nucleon observable is also in agreement with the recent study of ref.\cite{morfouace2019}. It can be qualitatively understood from the fact that the free particles essentially come from the small rapidity region, which reduces the geometrical effect linked to the relative size of the overlap zone.  This very different behavior between
 $R_{QP}$ and  $R_{free}$, as well as the different quantitative values, underline the fact that it is very important that the same observable is used in the transport calculation as in the experimental data for a meaningful comparison.

Concerning the projectile energy dependence, the increase of $R$ for both observables reflects the shorter interaction time, increasing importance of nucleon-nucleon collisions and decreasing influence of the mean field, suggesting the importance of low energy experiments for precise measurements of the EoS properties.\\
\indent
In order to quantify the sensitivity of the isospin transport ratio to the density dependence of symmetry energy, in Fig. \ref{Isospin_imbalance_ratio_sensitivity} the lowest order isovector parameters  $E_{sym}$, $L_{sym}$ and $K_{sym}$ are varied over the present uncertainty taken from Table-IV of ref. \cite{Margueron2018a}. These three parameters are tuned independently i.e. two of them are kept fixed at the reference Sly5 EoS values, whereas the third one is varied. In this energy domain, sub-saturation densities are probed and the symmetry energy is lower for lower values of $E_{sym}$ and $K_{sym}$, and higher value of $L_{sym}$. This leads to a decreased diffusion, and higher values of $R$.
This behavior is observed for both $R_{QP}$ (left panels of Fig. \ref{Isospin_imbalance_ratio_sensitivity}) and $R_{free}$ (right panels of Fig. \ref{Isospin_imbalance_ratio_sensitivity}) but the dependence is more clear when the QP is considered.
Though all low order empirical parameters clearly contribute to the isospin ratio, a higher sensitivity is observed to the $L_{sym}$ parameter. This can be explained by the fact that   the lowest order parameter $E_{sym}$ is the one which is best constrained by nuclear structure experiments, while densities relatively close to saturation are explored in the Fermi energy regime,
reducing the importance of the higher order parameter $K_{sym}$.\\
\begin{figure}
\begin{center}
\includegraphics[width=\columnwidth]{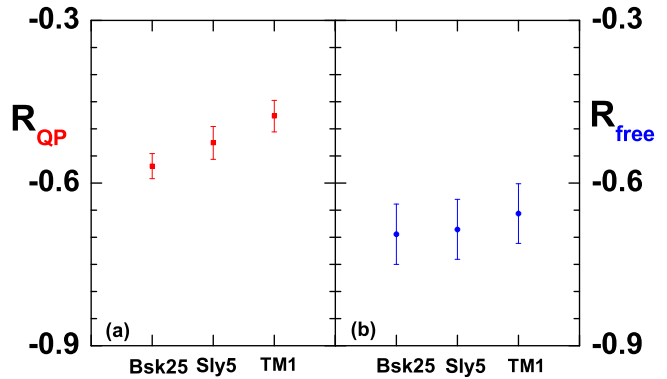}
\caption{Isospin transport ratio calculated from neutron to proton ratio of quasiprojectile (left panel) and free nucleons with $p_zc>0$ MeV (right panel) for $^{58}$Ni on $^{64}$Ni reaction at 52A MeV and 7 fm impact parameter calculated from Sly5, Bsk25 and TM1 equation of state.}
\label{Isospin_imbalance_ratio_BSK25_SLY5_TM1}
\end{center}
\end{figure}
\indent
The sensitivity analysis shows that we can potentially reduce the present uncertainties on EoS parameters with the $R_{QP}$ observable, but it does not allow producing quantitative predictions, because the different empirical parameters are correlated.
To study the discriminative power of isospin diffusion on realistic EoS, we have calculated the isospin transport ratio for the EoS  corresponding to three chosen popular functionals that cover the present uncertainty on the density dependence of the symmetry energy, namely the extended Skyrme
 Bsk25 \cite{Bsk25}, the standard Skyrme Sly5 \cite{Sly5} and the relativistic mean field parametrization TM1 \cite{TM1}.
The corresponding EoS parameters are reported in Table \ref{tab:tab1}.
We can see that those functionals especially differ on the $L_{sym}$ parameter, which shows the maximal sensitivity to the transport ratio. The results for the 52A MeV reaction  at $b$=7 $fm$ are reported in Fig.\ref{Isospin_imbalance_ratio_BSK25_SLY5_TM1}.
We can see that the different isospin transport predicted by the three functionals is within the error bars for the observable $R_{free}$. Conversely, a precise measurement of $R_{QP}$ should allow discriminating the different behaviors, thus contributing to the reduction of the present uncertainty of the isovector EoS.\\
\\
\begin{table}
\centering
\begin{tabular}{|c|c|c|c|}
\hline
Empirical & \multicolumn{3}{c|}{EoS} \\
\cline{2-4}
Parameter & BSK25 & Sly5 & TM1 \\
\hline
$E_{sat}$ &-16.03&-15.98& -16.26\\
$K_{sat}$ & 236& 230& 281\\
$Q_{sat}$ & -316& -364& -285\\
$Z_{sat}$ & 1354& 1592& 2014\\
$E_{sym}$ & 29.00& 32.03& 36.94\\
$L_{sym}$ & 36.9& 48.3& 111.0\\
$K_{sym}$ & -28& -112& 34\\
$Q_{sym}$ & 885& 501& -67\\
$Z_{sym}$ & -4919& -3087& -1546\\

\hline
\end{tabular}
\caption{Value in MeV of the lowest order empirical parameters of the three popular EoS sets used in this work. From top to bottom: saturation energy ($E_{sat}$), isoscalar incompressibility ($K_{sat}$), isoscalar skewness ($Q_{sat}$), isoscalar kurtosis ($Z_{sat}$), symmetry energy at saturation ($E_{sym}$), symmetry energy slope ($L_{sym}$), isovector incompressibility ($K_{sym}$), isoscalar skewness ($Q_{sym}$) and isoscalar kurtosis ($Z_{sym}$).}
\label{tab:tab1}
\end{table}
\section{Summary and conclusion}
The isospin diffusion  is studied for the $^{64,58}Ni$ on $^{64,58}Ni$ reactions in the Fermi energy domain, using the BUU@VECC-McGill transport model with a metamodelling for the nuclear equation of state.
It is observed that the isospin transport ratios obtained using the neutron to proton ratio of the projectile remnant, and the one of free particles emitted by it, are both independent of the secondary decay and sensitive to the density dependence of the symmetry energy. However, their absolute value and impact parameter behavior is different. This underlines the importance of using the same observable, when diffusion data are compared to transport model predictions. In particular, the $N/Z$ ratio of the quasi-projectile is a particularly robust observable for the transport models, and allows discriminating between the models for the EoS that cover the present uncertainty in the symmetry energy sector.
Precise measurements of this observable will be soon available\cite{manip}, and we expect that the comparison with the present calculations will be very helpful to progress further with respect to the present status of the knowledge of EoS.
\section{Acknowledgements}
This work was done under the support of “IFCPAR/CEFIPRA” Project No. 5804-3. S. Mallik acknowledges partial support from LPC Caen. We thank R. Bougault and D. Gruyer of LPC Caen and G. Chaudhuri of VECC for valuable discussions.

\end{document}